\documentclass[10pt]{article}

\usepackage{graphicx}
\begin{document}

\title{
Fluid-bicontinuous emulsions stabilized by colloidal particles
} 

\author{Paul S.~Clegg\footnotemark[2] \footnotemark[3] $^*$, 
Eva M.~Herzig\footnotemark[2] \footnotemark[3], Andrew B.~Schofield\footnotemark[2],\\ Stefan U.~Egelhaaf\footnotemark[4], 
Tommy S.~Horozov\footnotemark[5],
Bernard P.~Binks\footnotemark[5],\\ Micheal E.~Cates\footnotemark[2], and Wilson C.K.~Poon\footnotemark[2] \footnotemark[3]}%

\date{\today}

\maketitle

\footnotesize
\noindent \dag SUPA, School of Physics, University of Edinburgh, Edinburgh, EH9 3JZ, United Kingdom\\
\ddag Collaborative Optical Spectroscopy, Micromanipulation and Imaging Centre, 
University of Edinburgh, Edinburgh, EH9 3JZ, United Kingdom\\
\S Institut f\"{u}r Physik der kondensierten Materie, Heinrich-Heine-Universit\"{a}t, 
Universit\"{a}tstrasse 1, D-40225 D\"{u}sseldorf, Germany\\
\P Surfactant \& Colloid Group, Department of Chemistry, University of Hull, Hull, HU6 7RX, United Kingdom\\
\normalsize

\begin{abstract}
Using controlled spinodal decomposition, we have created a fluid-bicontinuous structure
stabilized by colloidal particles. We present confocal microscopy studies of these
structures and their variation with kinetic pathway. Our studies reveal a rigid multilayer
of colloidal particles jammed at the interface which prevents the liquids demixing for many
hours. The arrangement of two-fluid domains, interpenetrating on the meso-scale,
could be useful as a microreaction medium.
\end{abstract}

\vspace{0.5cm}

Emulsions are metastable soft materials of great industrial importance.
The liquid-liquid interface is conventionally stabilized by amphiphilic
surfactant molecules; the behavior of the resulting systems is well
understood~\cite{Bibette2002}. A less common variant, with the surfactants replaced by colloidal
particles or nanoparticles~\cite{Dinsmore2002, Aveyard2003, Lin2003}, 
was discovered over a century ago~\cite{Ramsden1903}. The colloids are much more
strongly pinned to the interface than are the surfactant molecules. Recent
research has begun to focus on the fundamental understanding of colloid
stabilized emulsions, which show strongly history-dependent 
properties~\cite{Aveyard2003, Arditty2004, Clegg2005}. This exemplifies a growing interest in
the behavior of materials that are irreversibly arrested far from equilibrium~\cite{Cates2000} 
(glasses, gels, etc.). 

The trapping
of solid particles at a liquid-liquid interface is governed 
primarily by the wetting angle $\theta_w$ of the liquid-liquid-solid contact 
line~\cite{Aveyard2003}.
This in turn depends on the surface chemistry (wettability) of the colloid, 
and on the distance from the critical point for liquid-liquid 
demixing~\cite{Winkelblech1908, LashMiller1908}, at which the liquid-liquid 
interfacial tension $\gamma$ vanishes. Maximal stabilization occurs close to the
neutral wetting condition ($\theta_w = 90^{\circ}$); 
a colloid of radius, say, $r = 0.25$~$\mu$m then reduces the
interfacial energy by~\cite{Aveyard2003}
$\Delta G_{int} = \pi r^2 \gamma (1 \pm
\cos(\theta_w)) \sim$ 10$^4$~k$_B$T (for a typical $\gamma \sim
1$~mN\,m$^{-1}$) so that adsorption to the interface cannot be reversed by
Brownian motion~\cite{Aveyard2003}. If such colloids are vigorously mixed 
in a pair of immiscible liquids, stable emulsion droplets
will form; the wetting angle and the relative phase volumes determine which 
liquid becomes the dispersed phase~\cite{BinksC2002}. 
As well as emulsification by conventional 
mixing~\cite{Yan1995, Dinsmore2002, Aveyard2003,
Lin2003}, dialysis~\cite{Giermanska-Kahn2002} and microfluidic emulsification~\cite{Subramaniam2005} 
also lead to droplets. 

\begin{figure}
\includegraphics[scale=0.7]{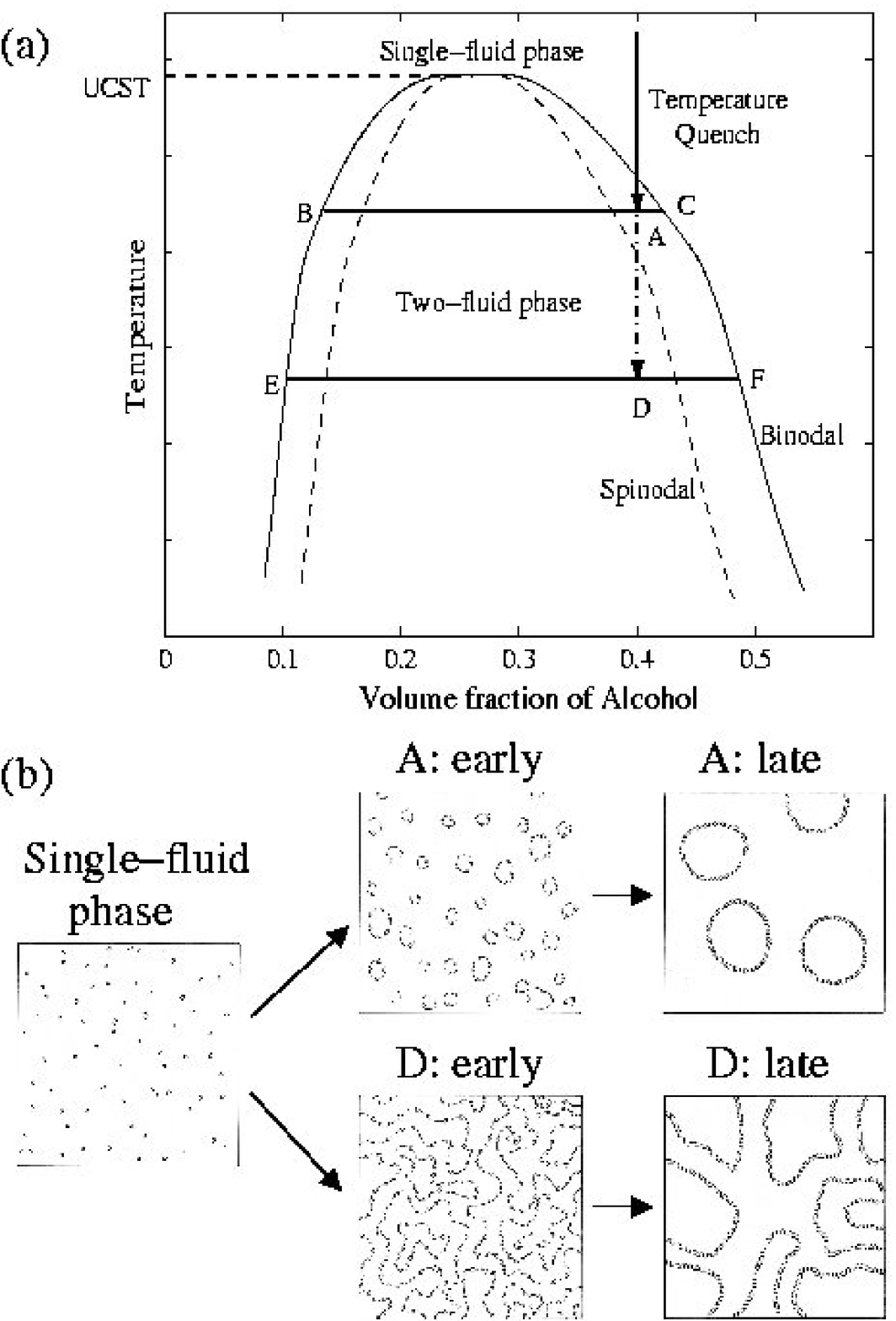}
\caption{\label{Phases_and_Kinetics} (a)
Generic alcohol-oil phase diagram showing binodal (solid
line) and spinodal (dashed line). The liquids are miscible above the
upper critical solution temperature (UCST). A shallow quench is pathway A
(top (b)): liquids separate via nucleation and a colloid-stabilized
droplet emulsion forms. The separated liquids have compositions
B and C. A deep quench is pathway D
(bottom (b)): liquids separate via spinodal decomposition leading to a
fluid-bicontinuous gel. The separated liquids have compositions E and F.}
\end{figure}

\begin{figure}
\includegraphics[scale=0.50]{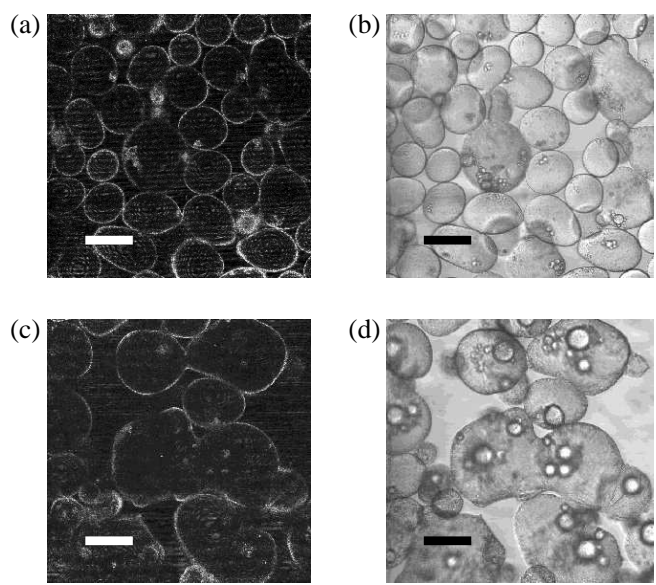}
\caption{\label{Pickering}
(a) Confocal and (b) bright-field microscopy images of a
hexane-in-methanol emulsion stabilized by silica colloids~\cite{recipe}. 
Fabrication corresponds to pathway A in Fig.~\ref{Phases_and_Kinetics}.
Confocal images show the interface
while bright-field images show the contents of the droplets.
(c) Confocal and (d) bright-field microscopy images of the same emulsion after
it has been reheated
then quenched to 0$^{\circ}$~C.
These images were all captured at room temperature
and the scale bars are 100~$\mu$m.}
\end{figure}

In a recent study~\cite{Clegg2005} 
we discovered a quite different route to particle-stabilized 
droplet emulsions based on binary-fluid phase 
separation. Silica
particles were initially dispersed in the single-fluid phase of alcohol-oil 
mixtures~\cite{recipe}. Following a
shallow quench, the liquids separate via nucleation
(Fig.~\ref{Phases_and_Kinetics}). The nuclei
coarsen until they are covered with a densely packed
layer of particles. Figs.~\ref{Pickering}a \& b
show the resulting droplets of hexane stabilized by a
multilayer coating of colloids. The colloid-laden interfaces
are rigid, with relatively high elastic moduli;
these structures are stable for at least a day. Under
certain conditions, droplets are aspherical with frozen
shapes~\cite{Clegg2005} (Fig.~\ref{Pickering}c \& d). 
This is the result of coalescence between
incompletely coated droplets. Some multiple emulsion (droplet within droplet) 
formation is also evident.

In this letter we report the use of this experimental approach to create a
quite different class of
structures. Specifically, we show that by
performing a deep quench at phase volumes close to 50:50 in colloid-doped
binary-fluid systems with a near-symmetric phase diagram
(Fig.~\ref{Phases_and_Kinetics}), bicontinuous
structures form, and are stable. In these bicontinuous
emulsions, as with the droplet emulsions formed by shallow 
quenches~\cite{Clegg2005}, the
liquid-liquid interface is coated by an immobile layer of colloidal
particles. With our protocols, this layer is several particles thick; similar
layers are seen in silica colloid-stabilized droplet emulsions~\cite{Binks2002}.
Though not yet conclusive (due primarily to the thin-slab geometry of our samples) 
our quenches promise the creation of a fully three-dimensional, bulk solid 
material simultaneously permeable to two immiscible solvents, with potential uses 
as a continuous-flow microreaction
medium~\cite{Patent,Stratford2005}.

Our experimental results confirm in broad outline, though not in every detail, a 
theoretical scenario for the formation of such materials presented by 
Stratford {\em et al.}~\cite{Stratford2005}, who dubbed them `Bijels' 
(bicontinuous interfacially jammed emulsion gels). These authors report computer 
simulations
that run for about 300~ns, showing sequestration of a monolayer of neutrally wetting 
nano-colloids, and drastic curtailment of
binary-fluid demixing, on that time scale. Extrapolating boldly by some nine decades 
in time, it was suggested in~\cite{Stratford2005} that Bijels should remain 
arrested, as metastable, three-dimensional bulk solids, on macroscopic timescales. 

The structures we have created strongly contrast with 
bicontinuous microemulsions~\cite{Andelman1987}, which are analogs in which
the particle layer is replaced by soluble surfactants.  The bicontinuous
surfactant layers continuously
break and reform; microemulsions are hence equilibrium structures and stable only
under a limited range of thermodynamic parameters. Our bicontinuous structures, stabilized
by interfacially jammed colloids, appear to be robust metastable states due presumably 
to the large energy barrier holding
the colloids in place.

The binary liquids used here are methanol-hexane (with upper critical solution temperature -- 
UCST 33.2$^{\circ}$~C for 27.5:72.5
alcohol:oil by volume) and ethanol-dodecane (with UCST 12.5$^{\circ}$~C for 36.1:63.9
alcohol:oil). The phase diagrams resemble Fig.~\ref{Phases_and_Kinetics}. For all
the experiments described in this letter a volume fraction half-way between critical and
50:50 was chosen (that is 39:61 methanol:hexane and 43:57
ethanol:dodecane). This is
a compromise between equal volumes and the critical composition, and, was chosen
to favor the formation of stable
bicontinuous structures. The silica colloids ($r = 0.220 \pm 0.007$~$\mu$m) were
synthesized using the St\"{o}ber procedure~\cite{Stober1968}. A single batch of colloids
was used for all of the
results presented here. The surface chemistry was modified
using the recipe described in ref.~\cite{Horozov2003}. For methanol-hexane,
10$^{-2}$~M of the silanizing agent (dichlorodimethylsilane - DCDMS) gave close
to neutral wetting (as estimated from the behaviour of macroscopic emulsion
droplets). For ethanol-dodecane, 10$^{-1}$~M of DCDMS was used. 

Prior to quenching, the liquids were
combined in an incubator (at 45$^{\circ}$~C for methanol-hexane and 28$^{\circ}$~C for
ethanol-dodecane). The colloids (2\% by volume) were dispersed
using an ultrasound processor (Sonics \& Materials) operating at 20~kHz for 2~minutes
at 2--3~W power. The sample cells, $0.2 \times 4.0 \times 50.0$~mm$^3$ (VitroCom),
were filled using capillary action. The filled cells were then quenched into the demixed
regime in various different ways. The end-point temperature of the quench was chosen to avoid 
freezing of either of the liquids.

\begin{figure}
\includegraphics[scale=0.42]{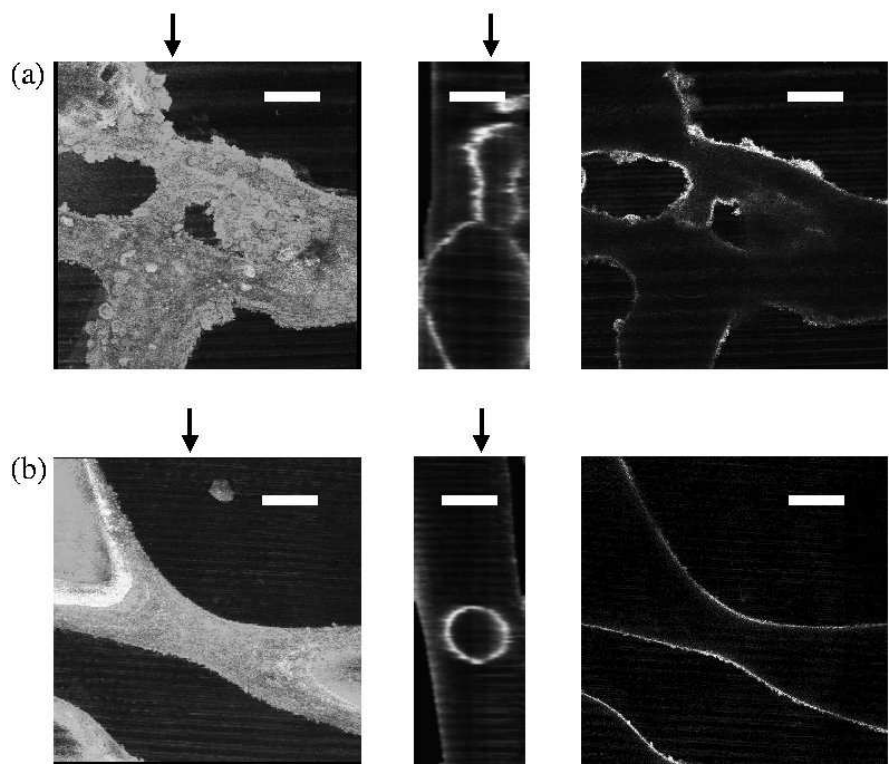}
\caption{\label{Convoluted} Left: Rendered confocal images of 3D structures viewed from
above; center: vertical slices corresponding to the position of the arrow on the left;
right: horizontal slice through the confocal stack on left (corresponding to the
position of the arrow in the center).
(a) Arrested bicontinuous structure. (b) An isolated fluid neck,
ending at contact lines with the 
glass plates confining the sample, stabilized by a colloidal multilayer. The creation of these
structures corresponds to pathway D in Fig.~\ref{Phases_and_Kinetics}.
All the images were recorded at
room temperature.
Scale bars 100~$\mu$m.}
\end{figure}

In Fig.~\ref{Convoluted} we present confocal microscopy images (rendered using the ImageJ
software package~\cite{ImageJ}) of bicontinuous 
structures created by deep quenching of a  methanol/hexane mixture
doped with colloidal silica particles
with near-neutral wetting.
The sample shown in Fig.~\ref{Convoluted}(a) was obtained
by first dipping it in liquid nitrogen for 5 seconds and then cooling it in a
CO$_2$ bath (-78$^{\circ}$~C) for 5 minutes. 
The sample shown in Fig.~\ref{Convoluted}(b) was obtained by cooling it in
liquid nitrogen for 30 seconds before submerging it in the CO$_2$ bath. The small
differences in fabrication procedure appears to be reflected in a smoother arrangement
of colloids on the interface. The images (Fig.~\ref{Convoluted}) 
are for a thin slab of material (thickness 200~$\mu$m) and show a 
particle-coated interface between the two fluid domains. This interface appears 
rough and lumpy, with the coating of particles about six colloids thick over 
much of its area (thicker in the lumpiest regions).
The structure is bicontinuous: there are connecting pathways of both fluids 
across both long and short dimensions of the cell. This 
requires the presence of free-standing fluid necks within the slab, and at 
least two of these are visible in Fig.~\ref{Convoluted}(a). These necks show 
clear departures  from the condition of constant mean curvature required for 
static mechanical equilibrium of a fluid-fluid interface (Fig.~\ref{Convoluted}(b)). 

Such variations in mean curvature are sustainable only by a rigidly frozen
interface, or a fluid film with near-zero interfacial tension (dominated by
curvature elasticity). In the latter case, the surface irregularities visible
in Fig.~\ref{Convoluted} would fluctuate in time; this is not seen.
We infer that the colloidal layer is rigid on 
the timescale of hours over which these observations were made. Were the 
interface to be fluid with finite tension, structures like the neck in 
Fig.~\ref{Convoluted}(b) would rapidly pinch off via the Rayleigh-Plateau 
instability~\cite{Stratford2005}.

Although our bicontinuous structures remain stable for more than a day, in some 
cases a slow evolution of the local structure is observable on this time scale. This could be 
due to residual `aging', as the structure lowers its energy locally by slight 
rearrangement~\cite{Stratford2005}. Alternatively, it could mean that the 
colloidal layer is not fully solid, but instead a viscoelastic fluid film. This 
would allow coarsening of the structure to slowly proceed indefinitely, in 
contrast to the arguments of Stratford {\em et al.}~\cite{Stratford2005}, who 
suggested that the interfacial stresses should cause a colloidal monolayer to 
undergo a glass transition into an amorphous solid. The amorphous interfacial films we 
observe are thicker than a monolayer, 
distributing these stresses over additional particles and possibly reducing the 
propensity to jam the layer fully. 

\begin{figure}
\includegraphics[scale=0.5]{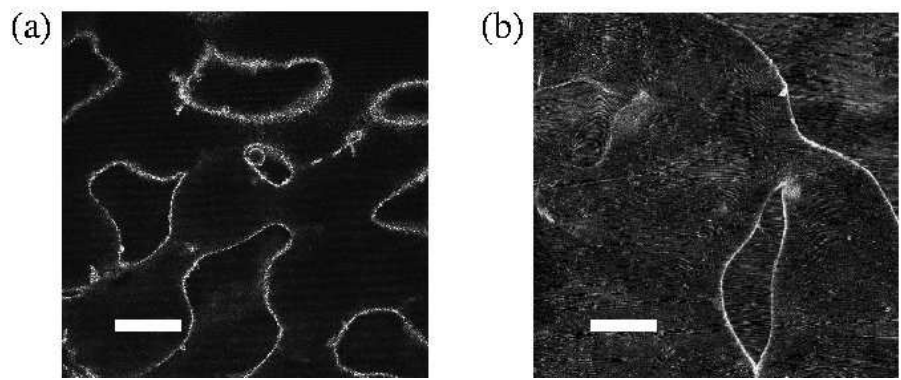}
\caption{\label{Length_Scale}
(a) Confocal image recorded at 0$^{\circ}$~C (using a cold
stage, Linkam Scientific Instruments) through an
ethanol-dodecane structure
stabilized by $\phi_v = 0.02$ silica colloids (preparation described in the text).
The characteristic separation between interfaces is $\xi \sim
150$~$\mu$m. (b) Confocal image recorded at room temperature of a
methanol-hexane structure
stabilized by $\phi_v = 0.01$ silica colloids~\cite{recipe}.
The characteristic separation between interfaces is $\xi \sim
300$~$\mu$m. Scale bars 100~$\mu$m.}
\end{figure}

Our work shows that, at least for the materials and sample geometries studied here, 
a bicontinuous interfacial structure indeed forms, and remains arrested for periods 
of hours or days. As detailed above,
we have found stable bicontinuous structures in two distinct pairs of binary 
liquids, requiring different surface treatment of the colloids and different
quenching schedules. 
Our results can be
combined with those of our previous study of shallow quenches~\cite{Clegg2005} 
(Fig.~\ref{Pickering}a \& b) to illuminate the role of the kinetic pathway in structure 
formation.
In droplet emulsions,
coalescence usually proceeds at a diameter-dependent rate until kinetic
arrest occurs on complete surface coverage~\cite{Arditty2003}.
In our bicontinuous quenches we cannot resolve the coarsening dynamics, but 
the final length scale of the arrested state should
be dictated by the colloid volume fraction $\phi_v$ and the thickness of the 
stabilizing layer.
For a layer $n$ colloids thick, the typical distance between interfaces~\cite{Stratford2005} 
is $\xi \sim  n r / \phi_v$. This scaling, which should apply also in the slab 
geometry, is broadly confirmed by varying $\phi_v$
(Fig.~\ref{Length_Scale}). (Droplet emulsions permit a broader size distribution which
can also depend on interfacial curvature; no simple scaling is expected in that case.)
The sample shown in Fig.~\ref{Length_Scale}(a) was quenched by submerging
it in liquid nitrogen for 30 seconds and then in an ice bath for 6 minutes. 
The deep quenches we employ here seem to be essential in promoting fluid 
bicontinuity~\cite{Bray1994}, 
which is not seen in emulsification by mixing \cite{Aveyard2003} even 
very close to $\theta_w = 90^\circ$.
Although shallow or slow quenches favor conventional droplet emulsions, warming 
such an emulsion back toward the miscible region of the binary fluid
phase diagram (Fig.~\ref{Phases_and_Kinetics}) results in an enhanced tendency 
for droplets to coalesce. Other studies~\cite{Clegg2005} show that the colloid-stabilized
droplets can remain intact well above the phase boundary.
Such warming, followed by a second quench, induces the formation of
fused droplets (Fig.~\ref{Pickering}c \& d). 
The cusps formed at the junction between these
droplets appear stable for long periods of time, again signifying solidification 
of the colloid film. Controlled coalescence of such droplets may offer an
alternative route to the creation of bicontinuous structures.

Our fluid-bicontinuous structures exemplify a
more general phenomenon. Formation of many composite soft
materials~\cite{Crawford1996, Meeker2000, Luo2003, Balazs2000, Hashimoto1993,
Tanaka1994, Chung2005} involve the co-evolution of a host system that undergoes a phase 
transition (e.g.,
liquid crystals~\cite{Crawford1996, Meeker2000}, binary
fluids~\cite{Luo2003, Balazs2000}, polymer blends~\cite{Hashimoto1993, Tanaka1994,
Chung2005}) 
with a dispersed constituent that alters or arrests that process 
(e.g., colloids~\cite{Meeker2000, Tanaka1994, Chung2005, Balazs2000},
polymers~\cite{Crawford1996, Luo2003}, block
copolymers~\cite{Hashimoto1993}). On cooling, the
host fluid begins to demix or form an ordered phase. The dispersed component 
then segregates either within a preferred phase or at the
interface. This segregation requires partial phase ordering of
the host, but then prevents this from going to completion. In our case, the true 
equilibrium state comprises full macroscopic phase separation (even with the 
particles present~\cite{Stratford2005}); but this state cannot be reached due to 
quasi-irreversible trapping of colloids at the liquid-liquid interface. This 
pathway has similarities to hardening of metals by impurities trapped at domain 
boundaries. It also resembles the formation of colloid-nematic 
gels (although these are not bicontinuous)~\cite{Meeker2000}, and might result 
in equally surprising bulk rheological properties~\cite{Patent}.

Beyond possible applications as microreaction media or rheologically active 
materials~\cite{Patent}, there are fundamental reasons to further explore the 
properties of our fluid-bicontinuous emulsions. First, changes
to the phase transition kinetics due to the colloidal
particles may arise: nonuniversal behaviour was
found for phase-separating polymer
blends with surfactant~\cite{Hashimoto1993}.
Second, the colloidal
films themselves represent interesting low-dimensional 
systems~\cite{Bausch2003, Lipowsky2005} whose glass-transition physics is coupled 
to a non-Euclidian curved space. 
Finally, alternative routes to the formation of arrested structures might involve 
pressure quenches, controlled coalescence, or highly energetic mixing.
These could enlarge the range of length scales and of fluid pairs for which 
bicontinuity is possible. 

We are very grateful to D.~Roux for illuminating
discussions. Funding in Edinburgh was provided by EPSRC Grant
GR/S10377/01.

\pagebreak

\end{document}